      [1999/12/01 v1.4c Il Nuovo Cimento]
\documentclass[nocopyright]{mycimento}

\usepackage{graphicx}  
\usepackage{slashed}  
\title{Higgs at LHC}
\author{S.~Bolognesi\from{ins:x},
  G.~Bozzi\from{ins:z} \atque
A.~Di~Simone\from{ins:y}}

\instlist{
\inst{ins:x} Universit\`a di Torino and INFN Sezione di Torino, 
  Via P. Giuria 1, I-10125 TORINO, Italy
  \inst{ins:z} Institut f\"ur Theoretische Physik, Universit\"at
  Karlsruhe, P.O.Box 6980, D-76128 Karlsruhe, Germany
  \inst{ins:y} INFN Sezione di Tor Vergata, Via Ricerca Scientifica 1, 
  I-00133 Roma, Italy}

\PACSes{\PACSit{14.80.Bn}{Standard Model Higgs Boson}}

\begin{document}

\maketitle

\begin{abstract}
An overview of recent theoretical results on the Higgs boson and its
discovery strategy at ATLAS~\cite{ATLASPTDR} and CMS~\cite{CMSPTDR} will
be presented, focusing on the main Higgs analysis effective with low
integrated luminosity ($<30$ fb$^{-1}$).
\end{abstract}

\section{Introduction}

One of the main tasks of the Large Hadron Collider (LHC) will be the
search for the Higgs particle~\cite{Gunion:1989we}, which is
responsible for the electroweak symmetry breaking of the Standard
Model (SM). A lower limit $m_H > $114 GeV was put on the Higgs mass by the non-observation of the so-called ''Higgs-strahlung'' process $e^+e^- \to HZ$ at LEP~\cite{Barate:2003sz}. Since radiative corrections to electroweak observables vary with $m_H$, a global $\chi^2$-fit of high-precision electroweak measurements performed at lepton and hadron colliders allows an indirect measure of the Higgs mass: the upper limit $m_H < $ 160 GeV has been obtained at 95\% confidence level ($\Delta\chi^2$=2.7)\cite{EWfit}.

A considerable effort has been devoted in recent years to improve the theoretical predictions both for the production mechanisms and the main background processes, hopefully leading to an overall improvement of the search strategies at the LHC.

This brief review is intended to summarize the status of the QCD corrections to Higgs boson production and decay at the LHC, and to present the main discovery strategies at ATLAS and CMS.

\section{Higgs production at the LHC}

\begin{figure}
\includegraphics[angle=-90,width=7.5cm,viewport=500 73 553 700]{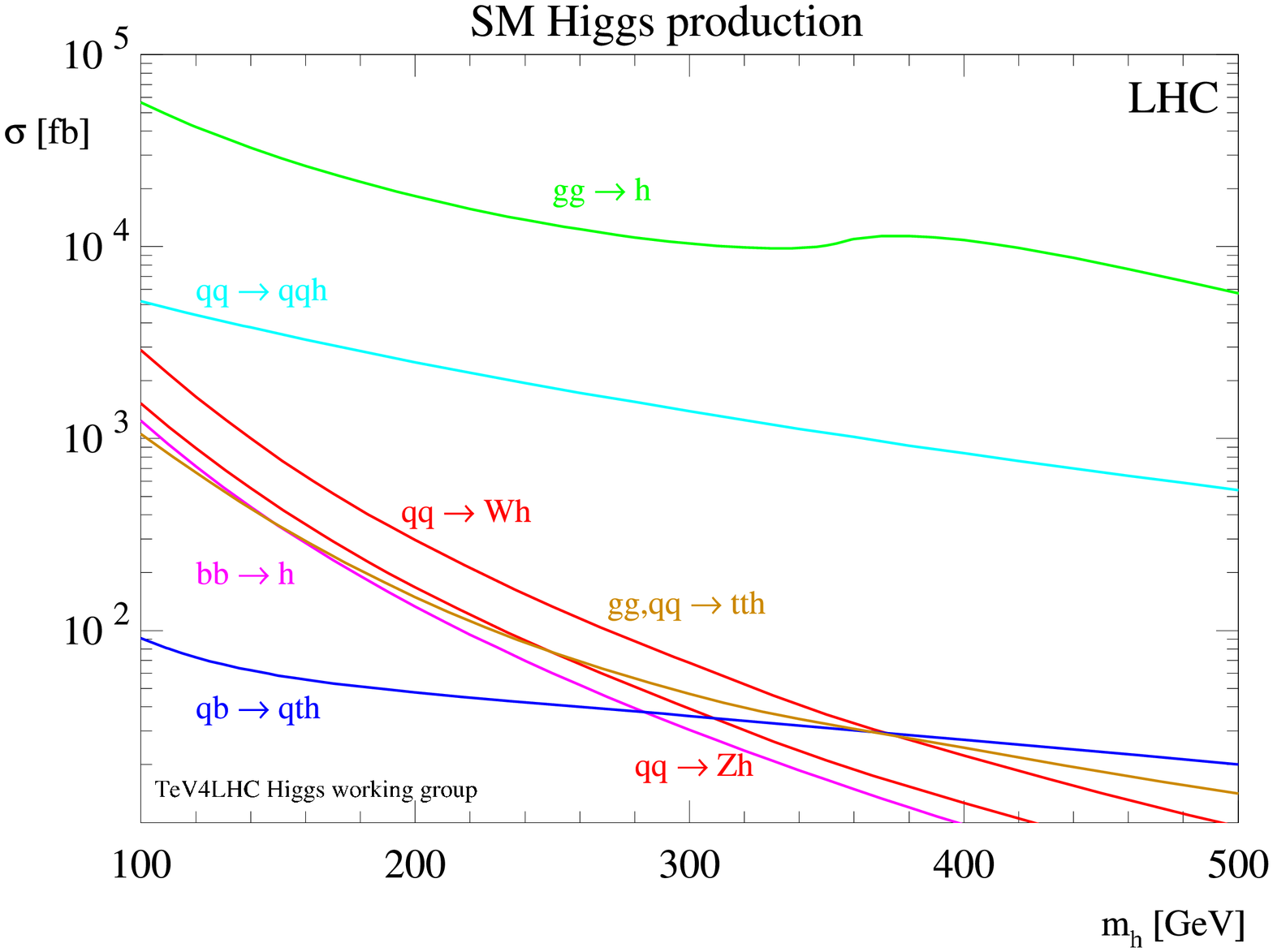}
\includegraphics[height=6.5cm, width=6cm]{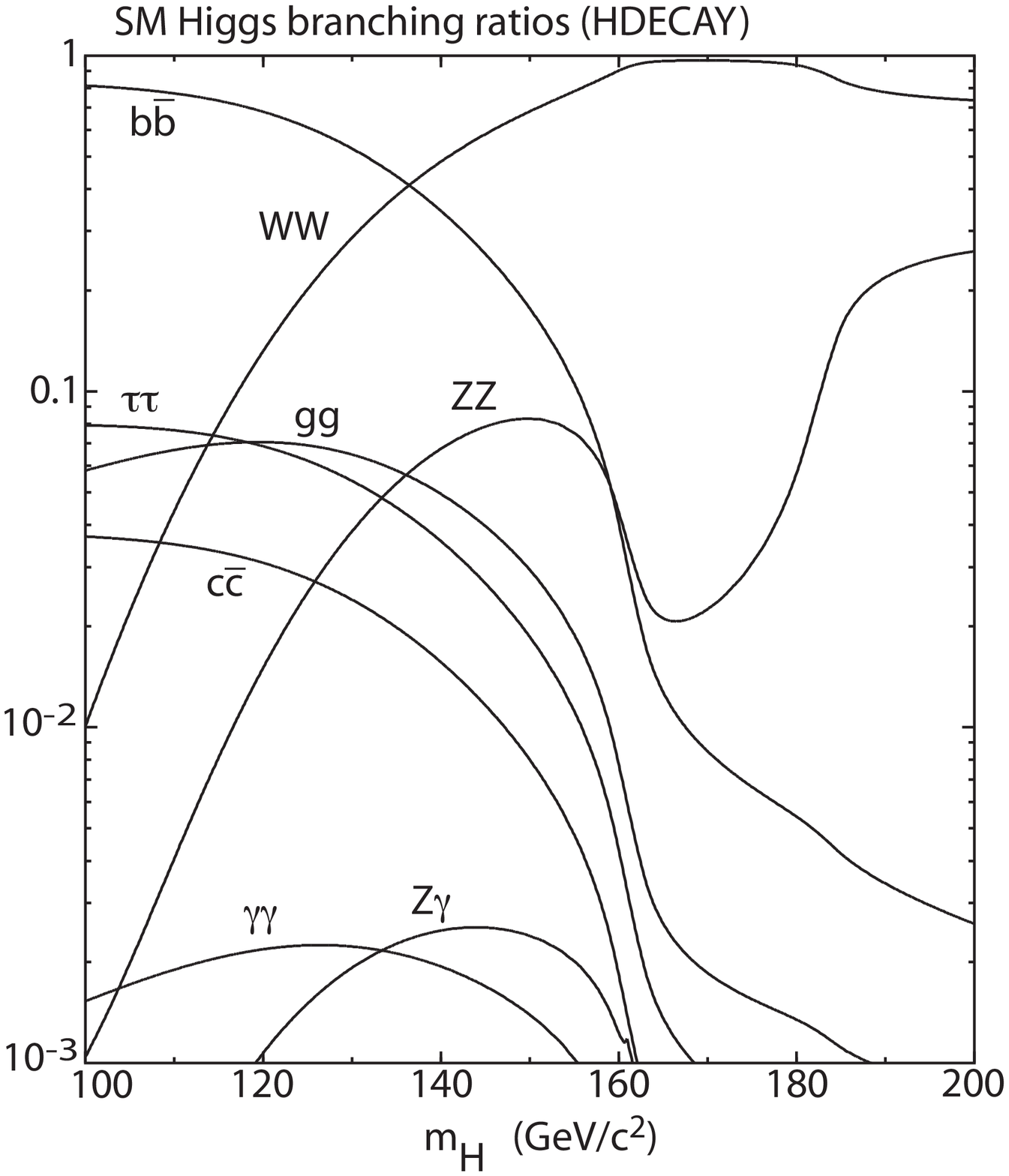}
\caption{Higgs production cross sections (left) and branching ratios
  (right) at the LHC (from \cite{Hahn:2006my}) as function of Higgs mass.}
\label{Hprod}
\end{figure}

In fig.~\ref{Hprod} (left), taken from \cite{Hahn:2006my}, the relevant cross
sections for Higgs production at the LHC are shown as a function of the Higgs mass. The results refer to fully inclusive cross sections and no acceptance cuts or branching ratios are applied. In this section, we describe the state-of-the-art of theoretical calculations for the main production channels.

\subsection{Gluon fusion} 

At the LHC, mainly because of the large gluon luminosity, the dominant production channel over the entire mass range will be the gluon fusion process $gg \rightarrow H$, where the Higgs couples to gluons through a heavy-quark loop. The total cross section at leading-order (LO) in QCD perturbation theory (${\cal O}(\alpha_{s}^{2})$) was computed more than 30 years ago \cite{Georgi:1977gs}. The next-to-leading order (NLO) corrections \cite{Dawson:1991zj, Djouadi:1991tk, Spira:1995rr} yield a K-factor of about 80-100\%, thus making a NNLO calculation explicitly needed. The complexity involved with the heavy-quark loop makes the computation of higher-order corrections extremely difficult. Considerable simplifications arise in the large-$m_{t}$ limit ($m_{t} \gg m_{H}$), where it is possible to introduce an effective lagrangian \cite{Ellis:1975ap} directly coupling the Higgs to gluons:
\begin{equation}
\label{eff}
{\cal L}_{\rm eff} =  -\frac{1}{4} [1-\frac{\alpha_{s}}{3\pi}\frac{H}{v}(1+{\rm \Delta}) ] \ {\rm Tr} \ {\cal G}_{\mu\nu} {\cal G}^{\mu\nu} ,
\end{equation}

where the coefficient $\rm \Delta$ is known up to ${\cal O}(\alpha_{s}^{3})$ \cite{Chetyrkin:1997iv}. It was shown~\cite{Kramer:1996iq} that NLO calculations based on the effective lagrangian approximate the full NLO result within 10\% up to $m_{H}$=1 TeV. The reason for the high accuracy of this approximation is the fact that the Higgs particle is predominantly produced in association with partons of relatively low transverse-momenta, which are unable to resolve the heavy-quark loop \cite{Catani:2001ic}.
The next-to-next-to-leading order (NNLO) corrections have been
evaluated in the large-$m_{t}$ limit
\cite{Catani:2001ic,Harlander:2000mg,Harlander:2001is,NNLOtotal}. In
the case of a light Higgs boson ($m_{H}\sim$ 100-200 GeV), the K-factor with respect to NLO is about 10-25\% and the scale dependence is reduced to 10-15\%, thus improving the convergence of the perturbative series. Higher-order perturbative contributions can be reliably estimated by resumming multiple soft-gluon emission: the resummation program has been carried out at the full next-to-next-to-logarithmic (NNLL) level \cite{Catani:2003zt} and at the ${\rm N^{3}LL}$ level \cite{Moch:2005ky}, providing a further 7-8\% increase w.r.t. NNLO and reducing scale dependence to less than 4\%. Also the NLO EW contributions have been computed~\cite{Aglietti:2004nj}, showing a 5-8\% effect below the $m_{H} = 2 m_{W}$ threshold. 

Realistic experimental analysis, including exact kinematics of the final states, are only possible if reliable theoretical predictions for the Higgs differential ($q_{T}$ and $y$) distributions are available. The most advanced predictions at present are the NNLO fully exclusive distribution \cite{Anastasiou:2005qj} and the parton level event generator (including Higgs decays) \texttt{HNNLO} \cite{Catani:2007vq}. In the region of small transverse-momentum, in addition to these fixed-order results, $q_{T}$-resummation has been performed at the NNLL level with inclusion of the rapidity dependence\cite{Bozzi:2003jy}, while joint (threshold and $q_{T}$) resummation has been performed at the full NLL level \cite{Kulesza:2003wn}, leading to very precise predictions and to an overall excellent convergence of the perturbative result.

\subsection{Vector boson fusion (VBF)} 

This production mechanism occurs as the scattering between two
(anti)quarks with weak boson ($W$ or $Z$) exchange in the t-channel
and with the Higgs boson radiated off the weak-boson propagator. Even
though the Higgs VBF production cross section is somewhat smaller
($\sim$ 20\%) than the gluon fusion one, several phenomenological features make VBF a very promising channel for the LHC: 
\begin{itemize}
\item since the parton distribution functions (pdfs) of the incoming
valence quarks peak at values of the momentum fractions $x\sim$ 0.1 to 0.2, this process tends to produce two highly-energetic outgoing quarks;
\item the large weak boson mass provides a natural cutoff on its
propagator: as a consequence, the jets from the two outgoing quarks are produced with a transverse energy of the order of a fraction of the weak boson mass and thus with a large rapidity interval between them (typically one at forward and the other at backward rapidity);
\item since the exchanged weak boson is colourless, no further hadronic production occurs in the rapidity interval between the quark jets (except for the Higgs decay products).
\end{itemize}
The LO partonic cross section can be found in \cite{Cahn:1983ip}. Gluon radiation can only occur as bremsstrahlung off the quark legs: NLO corrections to Higgs production via VBF have been computed for the total cross section~\cite{Han:1991ia} and for Higgs production in association with two jets~\cite{Figy:2003nv}. They have been found to be typically modest (5-10\%) and the scale uncertainty is at the percent level, mainly because of the good precision to which the valence quark pdfs in the intermediate-x regions are known.

\subsection{Associated production with top} 

The Higgs boson is radiated off one of the two tops in the $q\bar q,
gg$ s-channel or off the top propagator in the $gg$ t-channel at
LO. This channel can be important in the low-mass region (provided a
good $b$-tagging and a high luminosity are reached), where it allows to
search for $H \to b\bar b$ decay and can be useful to measure the
$t\bar t H$ Yukawa coupling. The QCD corrections to the LO cross
section (computed in \cite{Kunszt:1984ri}) involve the computation of
massive pentagons and enhance the cross section by $\sim$ 20\%, with a
residual scale dependence of $\sim$ 15\%~\cite{Beenakker:2002nc}.

\section{Low Higgs mass searches}
In fig.~\ref{Hprod} (right) the branching ratios are shown as function of the
Higgs mass for the low-intermediate Higgs mass region 100 GeV $< m_H
<$ 200 GeV, favourite by the electroweak precision
measurements~\cite{EWfit}. Unfortunately, the low Higgs mass
region is the most challenging 
for the Higgs detection because the biggest Higgs
branching ratios (BR) are into heavy quarks or $\tau$
leptons~\cite{branchRatios}, 
which are difficult to be disentangled from the huge QCD background.\\
For very low Higgs mass ($m_H \sim$~120 GeV) the $H \rightarrow
b\bar{b}$ decay channel (BR $\sim 70\%$) is exploited in the
associated Higgs production 
with $t\bar{t}$. The request of two additional top quarks helps to cut
the huge amount of $b\bar{b}$ QCD background but it makes the final
state very complex ($bbbbWW$).\\ 
For Higgs mass up to 130 GeV the most promising decays are into
photons and $\tau$ leptons. Both the channels have high
background rate due to fakes, therefore they are also studied in the
VBF production of the Higgs to achieve a better
significance. The decay $H \rightarrow \gamma\gamma$ is known up to 3-loop QCD \cite{Steinhauser:1996wy}, while the irreducible $pp \rightarrow \gamma\gamma$ background is available at NLO in the program DIPHOX \cite{Binoth:1999qq}, which also includes all relevant photon fragmentation effects. The loop-mediated process $gg \rightarrow \gamma\gamma$ contributes about 30\% to the background and has been calculated at ${\cal O}(\alpha_{s}^{3})$ \cite{Bern:2002jx}.

\subsection{$t\bar{t}H \rightarrow t\bar{t}b\bar{b}$}
ATLAS and CMS study this channel in all the combinations of the $W$ decays:
$b\bar{b}b\bar{b}l\nu l\nu$ with $\sigma \sim$~0.02~pb, $b\bar{b}b\bar{b}l\nu jj$ with $\sigma \sim$~0.10~pb,
$b\bar{b}b\bar{b}jjjj$  with $\sigma \sim$~0.20~pb, where $l$ stays for $e$ or
$\mu$. These final states involve many systematics
(\emph{e.g.}, effect of alignment on $b$-tagging, jet and missing energy calibration) 
which need to be measured from the data using dedicated
control samples. Moreover, in addition to the physics QCD
background ($t\bar{t}$+jets with $\sigma \sim$ 350 pb,
$t\bar{t}b\bar{b}$ with $\sigma \sim$ 3.3 pb), there is
a big amount of combinatorial background due to the crowding of these final states. 
Therefore multivariate analysis techniques
must be used to recognize the $b$-jets coming from the Higgs decay.

Both the experiments quote a quite low significance for the Higgs discovery
in this channel with 30 fb$^{-1}$: 
the most recent results are from CMS~\cite{ttH_CMSNOTE} and indicate a significance
smaller than 1 for the combined analysis, considering all the
systematics; the ATLAS analysis is quite old, it relays on a fast
simulation of the detector and it quotes a significance of about 2-3
for the semileptonic final state only~\cite{ttH_ATLASNOTE}.

\begin{figure}
\includegraphics*[height=5.5cm,viewport=19 16 190 140]{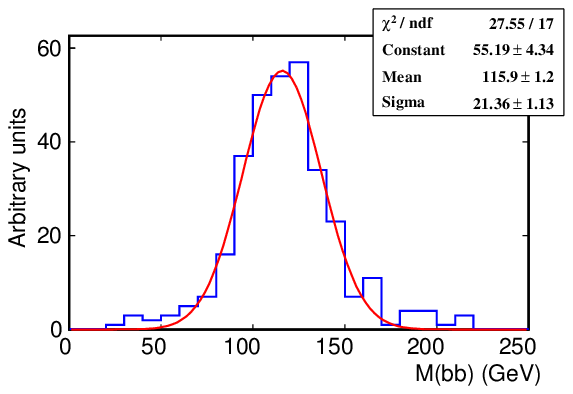}
\includegraphics[width=5.5cm]{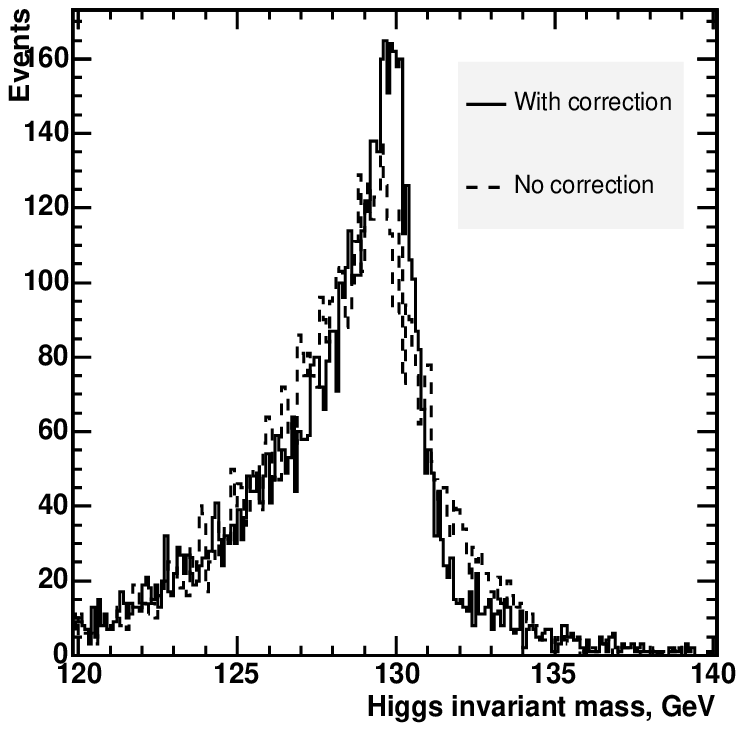}
\caption{Left: reconstructed Higgs mass in the channel $t\bar{t}H
\rightarrow b\bar{b}b\bar{b}l\nu l\nu$ in CMS. The jets are calibrated using
the PTDR II recommandation 1 settings and they are matched ($R<0.3$)
to the Monte Carlo $b$ partons from Higgs. 
Right: effect of the primary vertex reconstruction in CMS for
the $H\rightarrow\gamma\gamma$ channel. The dashed line shows the
reconstructed Higgs mass when no vertex correction is applied.}
\label{ttH_masses}
\end{figure}

However the analysis of this channel suffers of some drawbacks which can be solved
in the future. The low trigger efficiency (mainly in the fully hadronic
final state) can be raised with a dedicated, more complex trigger
menu. The $b$-tagging performances are optimal for jets with $p_T
\sim$~80~GeV, while this channel contains many low $p_T$ jets. On this
kind of jets also the reconstruction and calibration performances are
quite low, as shown in fig.~\ref{ttH_masses} (left). Both these aspects, the $b$-tagging and the jet measurement, 
can be improved exploiting a particle-flow approach to the jet reconstruction.

\subsection{$H \rightarrow \gamma\gamma$}
The branching ratio for this decay channel is actually very low (0.2$\%$ for $m_H=130$ GeV). On the other hand, the final state is very clean, thus allowing a bigger suppression of the backgrounds.
Reconstruction of the primary vertex is crucial for this analysis. ATLAS and CMS
have different approaches to the problem, which take into account the
different electromagnetic calorimetry used by the experiments.\\
ATLAS can use the longitudinal granularity of its EM calorimeter to
reconstruct the direction of the photons, thus achieving a precision
on the z coordinate of the primary vertex of $1.6$ cm, to be compared
with the $56$ mm spread coming from the LHC beam parameters. In
addition, high-$p_T$ tracks reconstructed in the inner detector can be 
included in the fit to improve the precision, leading to a precision
of about $40$ $\mu$m. \\
In CMS, where there is no longitudinal segmentation of the
calorimeter, only the reconstructed tracks can be used to fit the
primary vertex, and the resulting precision is $5$ mm (in a low
luminosity scenario). Fig.~\ref{ttH_masses} (right) shows the effect of vertex reconstruction on the reconstructed Higgs mass. If proper vertex position is calculated from the tracks, the number of events inside a window of $5$ GeV around the peak increases of about $15\%$.\\
Together with vertex reconstruction, $\gamma/\pi^0$ and $\gamma$/jet
discrimination play an important role in the analysis for this decay
channel, since they are crucial in rejecting the high reducible
background. The signal cross section is about $86$ fb at $m_H=130$ GeV, thus being three orders of magnitude below the cross section for $\gamma$+jet final state from
background processes, and a factor $10^6$ below the one for jet+jet final state. This calls for severe requirements in terms of jet and $\pi^0$ rejection.\\
CMS will use isolation contraints against jets, while $\pi^0$ rejection will be based on a neural network using as input several variables related to shower shapes, plus the information from pre-shower detectors in the endcaps.
An extensive use of the calorimeter transverse granularity, of hadronic leakage and shower shape parameters, will allow ATLAS to achieve a total rejection factor of about 3000 for $\gamma$ efficiency of $80\%$.\\
Given the amount of material in the inner trackers of the two experiments, photon conversions are not negligible in these studies, but must be recovered by means of dedicated reconstruction algorithms.\\
Signal significances can be improved using associated production studies, or more advanced analysis techniques (neural networks, likelihood, categories) and both ATLAS and CMS are exploring several possibilities.\\
Expected signal significances ($m_H=130$ GeV) at $30$ fb$^{-1}$ are 6.0 (cut based) and 8.2 (neural network) for CMS~\cite{gg_CMSNOTE} and 6.3 (cut based) for ATLAS.

\subsection{$H \rightarrow \tau\tau$}

\begin{figure}
\includegraphics[width=6.6cm]{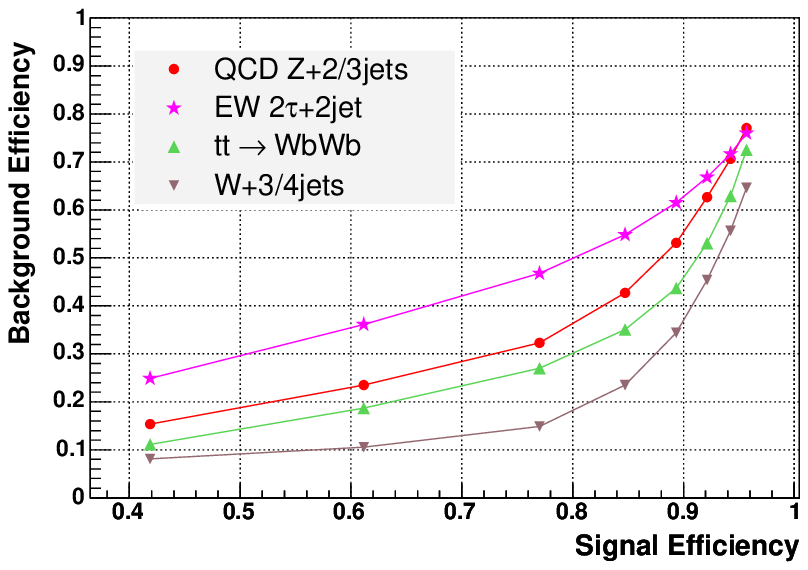}
\includegraphics[width=6.6cm]{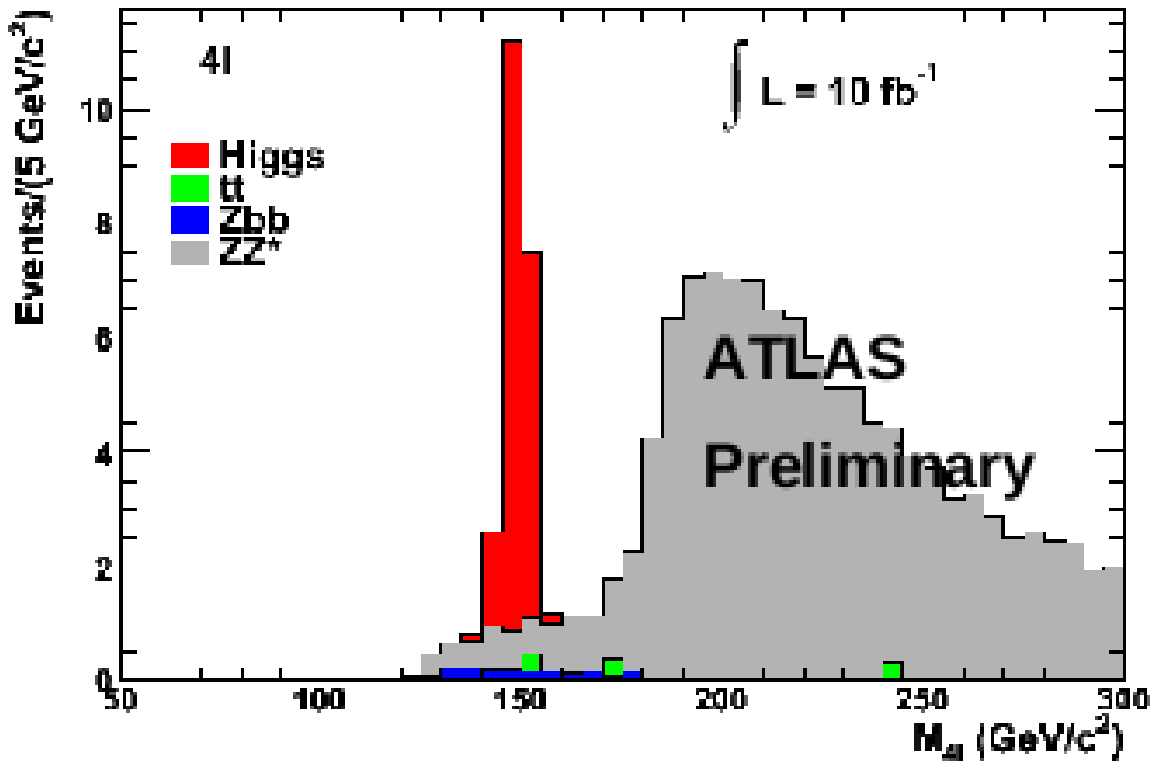}
\caption{Left: CMS results for central jet veto on the VBF $H\rightarrow\tau\tau$ 
channel. Background vs signal ($m_H=135$ GeV) efficiencies for different threshold values (10, 15, 20, ..., 45 GeV)~\cite{tautau_CMSNOTE}. Right: invariant mass of the two $Z$ bosons for the $H \rightarrow ZZ \rightarrow 4l$ channel as expected in the ATLAS experiment. Signal and main backgrounds are shown.}
\label{CJVHtt_HZZ}
\end{figure}

The high background rate for this final state makes it impossible to
study the $gg$ production channel. Both experiments are thus focusing
on VBF production where the additional jets in the final state allow
to improve significantly the signal over background ratio,
compensating for the smaller production cross section: $\sigma \sim 80$ fb at $m_H=135$ GeV. 
All possible final states (lepton-lepton, lepton-hadron, hadron-hadron) are presently under study in ATLAS, while CMS focused on the recent past only on the $lh$ decay channel.\\
The irreducible background for this decay channel comes from
$Zjj$ (QCD and electroweak) processes. In addition, several reducible backgrounds need to be taken
into account, such as QCD multijet, $W$+jet, $Z/\gamma$+jet,
$t\bar{t}$.\\
For leptonic and semi-leptonic final states, trigger menus based on single leptons and single leptons plus $\tau$s will be used, while for the fully hadronic decay channel, the most promising trigger configuration is $\tau$ plus missing $E_T$.\\
The VBF production channel allows for Forward Jet Tagging, where the event is searched through looking for the two additional quark initiated jets, which are typically located in opposite hemispheres and have high $p_T$ values. Both the experiments use similar techniques, looking for the two highest $p_T$ jets and requesting that they have opposite $\eta$ sign.\\
Another important characteristic of VBF events is the absence of jets
in the central rapidity region. This can be exploited implementing a
Central Jet Veto, which rejects events with jets in the central region
of the detector (apart from the jets identified as
$\tau$s). Fig.~\ref{CJVHtt_HZZ} (left) shows the background vs signal selection efficiencies for different values of the threshold applied on the jet energy in the central jet veto.\\
Expected signal significance ($m_H=130$ GeV) at $30$ fb$^{-1}$ is 4.4
($lh$ final state) and 5.7 ($lh$ plus $ll$) for ATLAS~\cite{tautau_ATLASNOTE} and 3.98 ($lh$
final state) for CMS~\cite{tautau_CMSNOTE}.

\section{$H \rightarrow VV$ channels}

In the high mass region ($m_H>150$ GeV) the most promising channels
are the ones with the Higgs decaying into two vector bosons ($WW$ or
$ZZ$). Their effectiveness of course follows very closely the shape of
the branching ratio curves. At about $160$ GeV the most interesting
channel is $WW$, while for heavier higgses, $ZZ$ becomes dominant once
the threshold for the on-shell production of the second $Z$ is approached. 
For $m_H>350$~GeV, the $t\bar{t}$ channel becomes
available, and the discovery potential for these channels is thus
reduced.\\
 The theoretical cross sections for these decay channels are known at 3-loop QCD for $H\to VV$ \cite{Kniehl:1995at} and 2-loop QCD for $H\to t\bar t$ \cite{Harlander:1997xa}. The backgrounds to the $H \to VV$ decay channel are also known at NLO QCD for $WW \to l\nu l\nu, ZZ \to 4l$ \cite{Campbell:1999ah} and for $VV$ production via VBF \cite{Bozzi:2007ur}.\\
Moreover, the VBF production channel ($VV \rightarrow VV$) is
interesting per se, since it is a powerful probe of the electroweak
symmetry breaking mechanism. In these processes, either the Higgs is
found in the s-channel, or unitarity is violated in
SM at the TeV scale and new physics must appear.

\subsection{$H \rightarrow ZZ \rightarrow 4l$}

These processes are very interesting over
a wide mass range, mainly for their very clean signature and quite
high production cross section ($\sim$ 38 fb$^{-1}$ at $m_H=135$ GeV). The most
critical region is 125-150 GeV, where one of the $Z$ bosons is
off-shell, leading to low-$p_T$ leptons which make the analysis more difficult.\\
The irreducible background $ZZ^*/\gamma^*\rightarrow 4l$
has a cross section of the order of tens of fb and it gives the biggest
contribution to background after analysis selection. In addition,
reducible background comes from $Zbb$ and $t\bar{t}$ processes, where
the needed rejection factors of $\sim 10^3$ and $\sim 10^5$ respectively
are achieved using lepton isolation and impact parameter cuts.\\
The crucial point of the analysis is lepton identification and reconstruction and actually the main systematic effects are expected to arise from lepton energy scale/resolution and lepton identification efficiency. In order to keep these effects under control, both ATLAS and CMS plan to measure lepton performance from data using $Z\rightarrow2l$ events.\\
Reconstructed invariant mass of the $ZZ$ pair is shown in
fig.~\ref{CJVHtt_HZZ} (right) for signal and the main backgrounds.

\subsection{$H \rightarrow WW \rightarrow l\nu l\nu$}
This fully leptonic final state ($\sigma \sim $ 0.5-2.5 pb)
is particularly clean but it has the
big drawback of not allowing the Higgs mass peak reconstruction. An
alternative variable to discriminate signal and backgrounds
is the azimuthal opening angle between the two
charged leptons ($\Delta\phi(ll)$). In the SM the Higgs boson has 0 spin so
the lepton (left-handed) and the anti-lepton (right-handed) tends to
go in the same direction and $\Delta\phi(ll)$ is small. This is a good assumption only for not too high Higgs
mass (below 200-250 GeV), otherwise the big boost of the $W$ bosons
pushes the two charged leptons into opposite directions.

Because of the absence of an Higgs peak, a careful strategy for background
normalization from real data is needed. The rate of the main backgrounds
($t\bar{t}$ with $\sigma \sim $ 86 pb, $WW$ with $\sigma
\sim$~12~pb, in the fully leptonic final states)
in the signal region is extrapolated from dedicated control regions,
using a rescaling factor evaluated from Monte Carlo.\\
A detailed study of the impact of theoretical uncertainties (mainly
due to the $gg\rightarrow WW$ description and double top with single top
interference) on this procedure has been carried out in the ATLAS
collaboration~\cite{HWW_ATLAS}, showing an uncertainty of about 5\% and 10\%
respectively on $WW$ and $t\bar{t}$ rate in the signal region. Given
these systematics, a Higgs discovery at $m_H \sim$ 160 GeV would
require less than 2 fb$^{-1}$.\\
A similar study from the CMS collaboration~\cite{HWW_CMSNOTE}
also takes into account the experimental
systematics (due to lepton identification, $b$-tagging, calorimetry
energy scale and jet energy scale) on the background normalization
procedure, showing that 2 fb$^{-1}$ (10 fb$^{-1}$) should be enough for a Higgs
discovery at 155~GeV~$<m_H<$~170~GeV (150~GeV~$<m_H<$~180~GeV).

The most recent progesses on this channel concern the strategies to measure
the interesting detector performances directly from data: the lepton
identification efficiency can be extrapolated from the efficiency
computed on single $Z$ sample exploiting the {\em tag-and-probe}
technique; to evaluate the impact of the $W$+jets background, the
lepton fake rate can be measured from QCD multi-jets events; finally,
the systematics on the missing energy can be estimated from the $W$ mass
measurement or by comparing $W$ and $Z$ with one lepton artificially removed.

\section{Higgs in the MSSM model}
\begin{figure}
\includegraphics[scale=0.85]{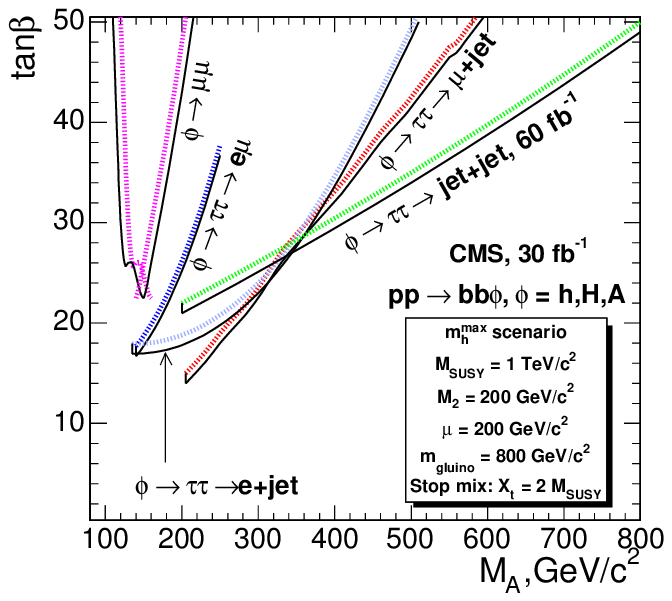}
\includegraphics[scale=0.85]{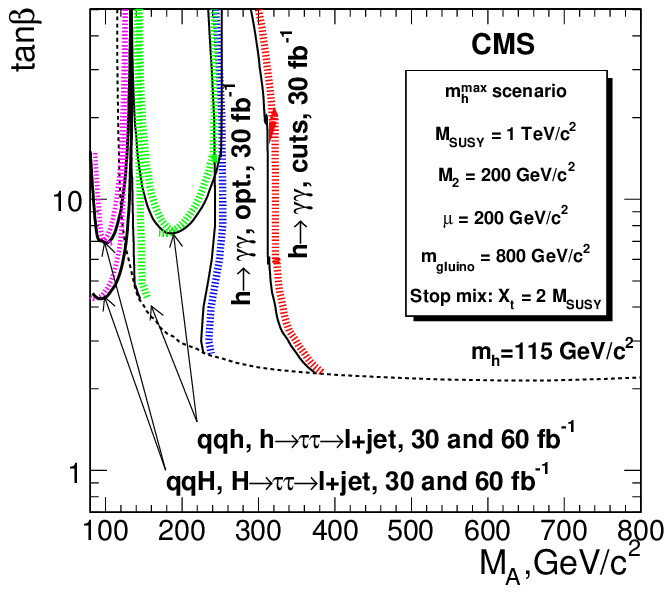}
\caption{Left: $5\sigma$ discovery regions in CMS with 30 fb$^{-1}$ for
the neutral Higgs bosons ($\phi = h,H,A$) produced in association with
$b$ quarks and decaying into  $\tau\tau$ and $\mu\mu$ in the $m_h^{max}$
scenario. Right: $5\sigma$ discovery regions in CMS with 30 fb$^{-1}$ for
the light neutral Higgs boson ($h$) decaying into $\gamma\gamma$ 
and for the light and heavy Higgs bosons ($h$ and $H$) produced in VBF
and decaying into $\tau\tau\rightarrow l+$jet in the $m_h^{max}$ scenario.}
\label{MSSM_CMS}
\end{figure}

The light neutral Higgs boson ($h$) has a similar behavior to the SM
Higgs, the most effective channel being $H \rightarrow \tau\tau$.\\
The heavier neutral Higgs bosons ($A$,$H$) are studied in different
channels, depending on the $\tan\beta$ value. For low $\tan\beta$, the
$gg$ fusion production is usually considered with peculiar
Higgs decay channels like $A \rightarrow Zh \rightarrow llbb$ or 
$A/H \rightarrow \chi^0_2\chi^0_2 \rightarrow 4l+\slashed{E}_T$. 
For high $\tan\beta$, the associated Higgs production with $b\bar{b}$
is studied with the Higgs decaying into $\tau\tau$ or $\mu\mu$ (with
quite low BR but much clean).
The $5\sigma$ discovery regions with 30 fb$^{-1}$ for the three neutral Higgs bosons at
CMS are shown in fig.~\ref{MSSM_CMS}.\\
Finally the charged Higgs bosons ($H^+,H^-$) are studied in different
channels depending on their mass. If $m_H<m_t$ then the most promising
channel is $tt\rightarrow tbH \rightarrow tb\tau\nu$. For $m_H>m_t$,
$gg\rightarrow tbH$ and $gb\rightarrow tH$ are the main production
mechanisms and $H\rightarrow tb$ and
$H\rightarrow\tau\nu$ (with lower BR but more clean) are the best
decay channels. All these final
states are very crowded, therefore they suffer from combinatorial
background in addition to big QCD background (mainly $t\bar{t}$+(b)jets
and $W$+(b)jets).

\section{Combined results}
\begin{figure}
\includegraphics[width=10cm]{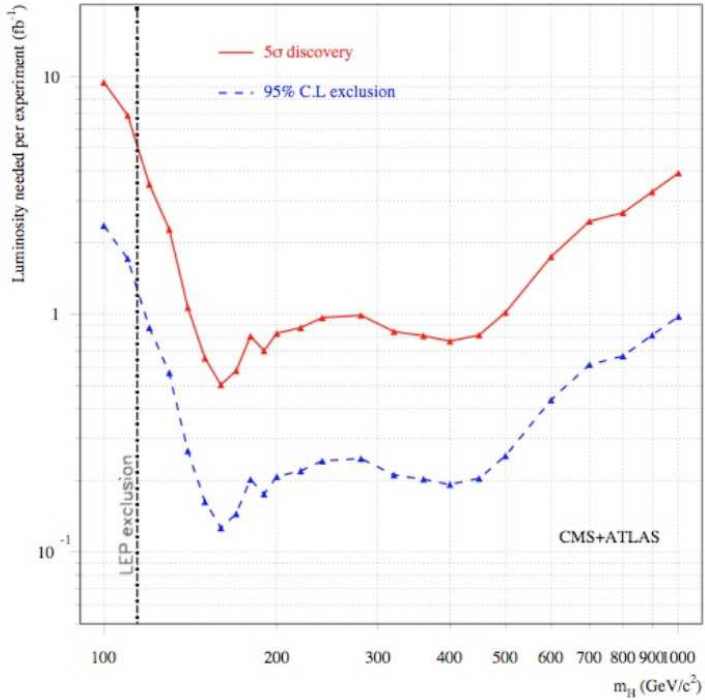}
\caption{Luminosity needed for the Higgs discovery and
exclusion at 95\% C.L. as a function of the Higgs mass, combining the
results from ATLAS and CMS.}
\label{combinedFig}
\end{figure}

In fig.~\ref{combinedFig} the combined results of the two experiments for the
Higgs discovery (or exclusion) are shown~\cite{combined}. The low Higgs mass region
is the most complex case because it requires the combination of several
channels ($H\rightarrow \tau\tau$, $H\rightarrow \gamma\gamma$ and
possibly $H\rightarrow bb$). The region 150 GeV $<m_H<$ 500 GeV is the
most favourable one, exploiting the clean channels $H\rightarrow ZZ
\rightarrow 4l$ in quite all this mass range except for $m_H\sim$ 160
GeV, where the $H\rightarrow WW\rightarrow l\nu l\nu$ decay channel dominates.

However, before a Higgs discovery can be claimed, some effort
will be necessary to understand the detector systematics (mainly regarding
jets, $\gamma$ fake rate, missing energy) and to perform a careful measurement of
the multi-jets background cross sections 
(like QCD jets, $V$+jets, $VV$+jets, $t\bar{t}$+jets, $b\bar{b}$+jets).

\end{document}